\documentclass[a4paper,11pt]{article}

\usepackage[english]{babel}
\usepackage[utf8x]{inputenc}
\usepackage[T1]{fontenc}
\usepackage[flushmargin]{footmisc}
\usepackage{setspace}
\usepackage[comma]{natbib}
\usepackage{float}
\usepackage{amsmath}
\usepackage{amsfonts}
\usepackage{amssymb}
\usepackage{ae}
\usepackage{caption}
\usepackage{tkz-graph}
\usetikzlibrary{shapes.geometric}

\usepackage[a4paper,left=2.5cm,right=2.5cm,top=3cm,bottom=3cm]{geometry}
\usepackage{graphicx}
\usepackage[colorinlistoftodos]{todonotes}
\usepackage[colorlinks=true, allcolors=blue]{hyperref}
\begin{document}

\begin{center}

{\LARGE An introduction to flexible methods for policy evaluation}

{\large \vspace{1.3cm}}

{\Large Martin Huber}\medskip

{\small {University of Fribourg, Dept.\ of Economics} \bigskip }
\end{center}

\bigskip

\noindent \textbf{Abstract:} {\small This chapter covers different approaches to policy evaluation for assessing the causal effect of a treatment or intervention on an outcome of interest. As an introduction to causal inference, the discussion starts with the experimental evaluation of a randomized treatment. It then reviews evaluation methods based on selection on observables (assuming a quasi-random treatment given observed covariates), instrumental variables (inducing a quasi-random shift in the treatment), difference-in-differences and changes-in-changes (exploiting changes in outcomes over time), as well as regression discontinuities and kinks (using changes in the treatment assignment at some threshold of a running variable). The chapter discusses methods particularly suited for data with many observations for a flexible (i.e.\ semi- or nonparametric) modeling of treatment effects, and/or many (i.e.\ high dimensional) observed covariates by applying machine learning to select and control for covariates in a data-driven way. This is not only useful for tackling confounding by controlling for instance for factors jointly affecting the treatment and the outcome, but also for learning effect heterogeneities across subgroups defined upon observable covariates and optimally targeting those groups for which the treatment is most effective.}

{\small \smallskip }



{\small \smallskip {\scriptsize The author is grateful to Colin Cameron, Selina Gangl, Michael Knaus, Henrika Langen, Michael Lechner, and Martin Spindler for their valuable comments. Address for correspondence: Martin Huber, University of Fribourg, Bd.\ de P\'{e}rolles 90, 1700 Fribourg, Switzerland; martin.huber@unifr.ch.
}\thispagestyle{empty}\pagebreak  }

{\small \renewcommand{\thefootnote}{\arabic{footnote}} %
\setcounter{footnote}{0}  \pagebreak \setcounter{footnote}{0} \pagebreak %
\setcounter{page}{1} }

\section{Introduction}\label{intro}

The last decades have witnessed important advancements in policy evaluation methods for assessing the causal effect of a treatment on an outcome of interest, which are particularly relevant in the context of data with many observations and/or observed covariates. Such advancements include the development or refinement of quasi-experimental evaluation techniques, estimators for flexible (i.e.\ semi- or nonparametric) treatment effect models, and machine learning algorithms for a data-driven control for covariates in order to tackle confounding, learn effect heterogeneities across subgroups and target groups for which the treatment is most effective.  Policy evaluation methods aim at assessing causal effects despite the problem that for any subject in the data, outcomes cannot be observed at the same time in the presence and absence of the treatment. As an illustration of this fundamental problem for causality, consider the treatment effect of a job application training for jobseekers on employment. Identifying this effect on the individual level requires comparing the employment state for a specific subject at a particular point in time with and without training participation. However, at a specific point in time, an individual can be observed to have either participated or not participated in the training, but not both. Therefore, treatment effects remain unidentified on the individual level without strong assumptions.

Formally, denote by $D$ a binary treatment, such that $D=1$ if for instance someone participates in a training and $D=0$ otherwise.  Furthermore, denote by $Y$ the observed outcome, e.g.\ employment. Following \cite{Rubin74}, let $Y(1)$ and $Y(0)$ denote the potential outcomes a subject would realize if $D$ was set to 1 and 0, respectively, e.g.\ the potential  employment state with and without training. It is assumed throughout that $Y(1)$ and $Y(0)$ only depend on the subject's own treatment and not on the treatment values of other subjects, which is known at the `Stable Unit Treatment Value Assumption', see \cite{Rubin1990}. Observed employment $Y$ corresponds to either $Y(1)$ if the individual receives the training ($D=1$) or to $Y(0)$ otherwise. The fact that not both potential outcomes are observed at the same time is formally expressed in the following equation:
\begin{eqnarray}\label{eq1}
Y= Y(1)\cdot D +  Y(0)\cdot (1-D).
\end{eqnarray}
It is easy to see that \eqref{eq1} is equivalent to $Y=Y(0)+D\cdot[Y(1)-Y(0)]$, where the observed outcome is the sum of the potential outcome without intervention and $D$ times $Y(1)-Y(0)$, i.e.\ the causal effect of $D$ on $Y$. As either $Y(1)$ or $Y(0)$ is unknown depending on the value of $D$, the treatment effect can in general not be identified for any subject.

Under specific assumptions, however, aggregate treatment effects are identified based on groups of individuals receiving and not receiving the treatment. Two parameters that have received substantial attention are the average treatment effect (ATE, denoted by $\Delta$) in the population, e.g.\ among all jobseekers, and the treatment effect on the treated population (ATET, denoted by $\Delta_{D=1}$), e.g.\ among training participants:
\begin{eqnarray}\label{ate}
\Delta=E[Y(1)-Y(0)],\quad \Delta_{D=1}=E[Y(1)-Y(0)|D=1].
\end{eqnarray}
One assumption yielding identification is statistical independence of treatment assignment and potential outcomes. Formally,
\begin{eqnarray}\label{random}
\{Y(1),Y(0)\}\bot D,
\end{eqnarray}
where `$\bot$' denotes statistical independence. \eqref{random} implies that there exist no variables jointly affecting the treatment and the potential outcomes. It is satisfied by design in experiments where the treatment is randomized, i.e.\ not a function of any observed or unobserved characteristics like education, gender, or income. The ATE is then identified by the mean difference in observed outcomes across treated and nontreated groups. This follows from the fact that by \eqref{eq1}, $E[Y|D=1]=E[Y(1)|D=1]$ and $E[Y|D=0]=E[Y(0)|D=0]$, while it follows from \eqref{random} that $E[Y(1)|D=1]=E[Y(1)]$ and $E[Y(0)|D=0]=E[Y(0)]$. As the average outcomes among treated and nontreated are representative for the respective mean potential outcomes under treatment and nontreatment in the population, $E[Y|D=1]-E[Y|D=0]=\Delta$.

When the treatment is not randomized, however, a mean comparison of treated and nontreated outcomes is generally biased due to selective treatment take-up, implying that subjects in the treated and nontreated groups differ in characteristics that also affect the outcome. Jobseekers attending a job application training could, for instance, on average have a different level of labor market experience or education than those not participating. Differences in the observed outcomes of treated and nontreated subjects therefore not exclusively reflect the treatment effect, but also the effects of such characteristics, which are thus confounders of the treatment-outcome relation. Formally, the selection biases for the ATE and ATET are given by
\begin{eqnarray}
E[Y|D=1]-E[Y|D=0]-\Delta&=&E[Y|D=1]-E[Y(1)]+E[Y(0)]-E[Y|D=0],\notag  \\
E[Y|D=1]-E[Y|D=0]-\Delta_{D=1}&=&E[Y(0)|D=1]-E[Y|D=0].
\end{eqnarray}

Different strategies have been developed for avoiding or tackling selection into treatment in order to identify causal effects. This chapter reviews the most prominent approaches, focusing on methods for flexible model selection and estimation particularly appropriate in big data contexts with many observations and/or variables. Section \ref{selobs} covers methods relying on selection-on-observables assumptions, implying that observed preselected covariates are sufficient to control for characteristics jointly affecting the treatment and the potential outcomes. Section \ref{prac} discusses practical issues to be verified in the data when invoking the selection-on-observables assumption, e.g.\ the similarity of treated and nontreaded subjects used for estimation in terms of observed characteristics,
as well as extensions e.g.\ to multivalued treatments and different treatment parameters. Section \ref{ML} covers causal machine learning, where observed covariates are not preselected, but it is assumed that important confounders can be controlled for in a data-driven way by machine learning algorithms. Section \ref{MLhet} outlines the application of machine learning for the data-driven detection of effect heterogeneities across subgroups defined upon observed covariates as well as for learning optimal policy rules to target subgroups in a way that maximizes the treatment effect.

Section \ref{IV} considers treatment evaluation based on instrumental variables. Here, treatment selection may be related to unobserved characteristics if a quasi-random instrument exists that affects the treatment, but not directly the outcome. Section \ref{did} discusses difference-in-differences methods, where identification hinges on common trends in mean potential outcomes under nontreatment over time across actually treated and nontreated groups. It also presents the changes-in-changes approach, which assumes that within treatment groups, the distribution of unobserved characteristics that affect the potential outcome under nontreatment remains constant over time. Section \ref{rdd} introduces the regression discontinuity design, which assumes the treatment probability to discontinuously change and be quasi-randomly assigned at a specific threshold value of an observed index variable. It also discusses the regression kink design, which assumes a kink in the (continuous) association of the treatment and the index variable at a specific threshold. Section \ref{conclusion} concludes.

\section{Selection on observables with preselected covariates}\label{selobs}

The selection-on-observables assumption, also called conditional independence or exogeneity, postulates that the
covariate information in the data is rich enough to control for characteristics jointly affecting the treatment and the outcome. This implies that one either directly observes those characteristics confounding the treatment-outcome relationship or that conditional on the observed information, the effects of unobserved confounders on either the treatment or the outcome (or both) are controlled for. As a further assumption known as common support, it is required that for any empirically feasible combination of observed covariates, both treated and nontreated subjects can be observed, which rules out that the covariates deterministically predict participation. Finally, the covariates must in general not be affected by the treatment, but measured at or prior to treatment assignment.

Denote by $X$ the vector of observed covariates and $X(1),X(0)$ the potential covariate values with and without treatment. Formally, the assumptions can be stated as
\begin{eqnarray}\label{assumpselobs}
\{Y(1),Y(0)\}\bot D|X, \quad 0<p(X)<1, \quad X(1)=X(0)=X,
\end{eqnarray}
where $p(X)=\Pr(D=1|X)$ is the conditional treatment probability, also known as propensity score. The first part of \eqref{assumpselobs} means that the distributions of the potential outcomes are conditionally independent of the treatment. This implies that $D$ is as good as randomly assigned among subjects with the same values in $X$. The second part says that the propensity score is larger than zero and smaller than one such that $D$ is not deterministic in $X$ and common support holds. The third part states that $X$ is not a function of $D$ and therefore must not contain (post-treatment) characteristics that are affected by the treatment, in order to not condition away part of the treatment effect of interest. This identification approach mimics the experimental context with the help of observed information. After creating groups with and without treatment that are comparable in the covariates, differences in the outcomes are assumed to be exclusively caused by the treatment.

The first part of \eqref{assumpselobs} is somewhat stronger than actually required for ATE identification and could be relaxed to conditional independence in the means (rather than all moments) of potential outcomes, $E[Y(d)|D=1,X]=E[Y(d)|D=0,X]$ for $d$ $\in$ $\{1,0\}$. In empirical applications it might, however, be hard to argue that conditional independence holds in means but not in other distributional features, which would for instance rule out mean independence for nonlinear (e.g.\ log) transformations of $Y$. Furthermore, the stronger conditional independence assumption in \eqref{assumpselobs} is required for the identification of distributional parameters like the quantile treatment effect, which corresponds to the effect at a particular rank of the potential outcome distribution. Also note that for the identification of treatment parameters among the treated (rather than the total) population like the ATET, \eqref{assumpselobs} can be relaxed to $Y(1)\bot D|X$, $p(X)<1$.

Let $\mu_d(x)=E[Y|D=d,X=x]$ denote the conditional mean outcome given $D$ corresponding to $d$ $\in \{1,0\}$ and $X$ equaling some value $x$ in its support. Analogous to identification under a random treatment discussed in Section \ref{intro}, $\mu_1(x)-\mu_0(x)$ under \eqref{assumpselobs} identifies the conditional average treatment effect (CATE) given $X$, denoted by $\Delta_x$:
\begin{eqnarray}\label{cate}
\Delta_x=E[Y(1)-Y(0)|X=x]=\mu_1(x)-\mu_0(x).
\end{eqnarray}
Averaging CATEs over $X$ in the population or among treated yields the ATE or ATET, respectively:
\begin{eqnarray}\label{condmeanselobs}
\Delta&=&E[\mu_1(X)-\mu_0(X)],\\
\Delta_{D=1}&=&E[\mu_1(X)-\mu_0(X)|D=1]=E[Y|D=1]-E[\mu_0(X)|D=1].\notag
\end{eqnarray}
Noting that the propensity score possesses the so-called balancing property, see \cite{rosenbaum1983}, such that conditioning on  $p(X)$ equalizes or balances the distribution of $X$ across treatment groups (i.e.\ $X \bot D | p(X)$), the effects are also identified when substituting control variables $X$ by $p(X)$:
\begin{eqnarray}\label{condmeanselobspscore}
\Delta&=&E[\mu_1(p(X))-\mu_0(p(X))],\\
\Delta_{D=1}&=&E[\mu_1(p(X))-\mu_0(p(X))|D=1]=E[Y|D=1]-E[\mu_0(p(X))|D=1].\notag
\end{eqnarray}

By basic probability theory, implying e.g.\ $\mu_1(X)=E[Y\cdot D|X]/p(X)$, and the law of iterated expectations, the ATE and ATET are also identified by inverse probability weighting (IPW), see \cite{Horvitz52}, using the propensity score:
\begin{eqnarray}\label{ipwselobs}
\Delta&=&E\left[\frac{Y\cdot D}{p(X)}-\frac{Y\cdot (1-D)}{1-p(X)}\right],\\
\Delta_{D=1}&=&E\left[\frac{Y\cdot D}{\Pr(D=1)}-\frac{Y\cdot (1-D)\cdot p(X)}{(1-p(X))\cdot\Pr(D=1)}\right].\notag
\end{eqnarray}
Finally, the effects follow from a combination of conditional mean outcomes and propensity scores based on so-called doubly robust identification using the efficient score function, see \cite{Robins+94}, \cite{RoRo95}, and  \cite{Ha98}:
\begin{eqnarray}\label{drselobs}
\Delta&=&E\left[\phi(X) \right],\textrm{ with }\phi(X)=\mu_1(X)-\mu_0(X)+\frac{(Y-\mu_1(X))\cdot D}{p(X)}-\frac{(Y-\mu_0(X))\cdot (1-D)}{1-p(X)},\notag\\
\Delta_{D=1}&=&E\left[\frac{(Y-\mu_0(X))\cdot D}{\Pr(D=1)}-\frac{(Y-\mu_0(X))\cdot (1-D)\cdot p(X)}{(1-p(X))\cdot\Pr(D=1)}\right].
\end{eqnarray}
Note that the identification results in \eqref{drselobs} coincide with those in \eqref{ipwselobs} and \eqref{condmeanselobs} because
\begin{eqnarray*}
&&E\left[\frac{(Y-\mu_1(X))\cdot D}{p(X)}-\frac{(Y-\mu_0(X))\cdot (1-D)}{1-p(X)}\right]=0\quad\textrm{ and}\\
&&E\left[\frac{-\mu_0(X)\cdot D}{\Pr(D=1)}-\frac{-\mu_0(X)\cdot (1-D)\cdot p(X)}{(1-p(X))\cdot\Pr(D=1)}\right]=E\left[\mu_0(X)\cdot\left(\frac{p(X)}{\Pr(D=1)}-\frac{p(X)}{\Pr(D=1)}\right)\right]=0.
\end{eqnarray*}

Assuming the availability of a randomly drawn sample, treatment effect estimation proceeds using the sample analogs of the identification results and plug-in estimates for $p(X), \mu_1(X), \mu_0(X)$ whenever required. When for instance considering the estimation of $\Delta_{D=1}$ based on \eqref{condmeanselobs}, an estimate of $\mu_0(X)$ for each treated observation is obtained as a weighted average of nontreated outcomes, where the weights depend on the similarity of the treated and nontreated observations in terms of $X$. One class of methods in this context are matching estimators, see for instance \cite{rosenbaum1983}, \cite{RosenbaumRubin1985}, \cite{Heck+98}, \cite{HeIcSmTo98}, \cite{DehejiaWahba99}, and \cite{LeMiWu11}. Pair matching, for instance, assigns a weight of 1 (or 100\%) to the most similar nontreated observation and of 0 to all others. $1:M$ matching estimates $\mu_0(X)$ based on the mean outcome of the $M$ most similar nontreated observations, where $M$ is an integer larger than 1. Radius or caliper matching defines a maximum tolerance of dissimilarity in $X$ and relies on the mean outcome of all nontreated observations within the tolerance. Compared to $1:M$ estimation, this may reduce the variance when many similar nontreated observations are available. Due to the multidimensionality of $X$, similarity is to be defined by a distance metric. Examples include the square root of the sum of squared differences in elements of $X$ across some treated and nontreated observation, either normalized by the inverse of the sample covariance matrix of $X$ (then called  Mahalanobis distance) or by the diagonal thereof (i.e.\ the variance). See \cite{Zh04} for a discussion of alternative distance metrics.

\cite{AbadieImbens01} show that in contrast to other treatment estimators, pair or $1:M$ matching does not necessarily converge with a rate of $n^{-1/2}$ to the true effect (i.e.\ is not $n^{-1/2}$-consistent) if $X$ contains citehan one continuous element, with $n$ being the sample size.  Second, even under $n^{-1/2}$-consistency, it does not attain the semiparametric efficiency bounds derived in \cite{Ha98}. Therefore, pair or $1:M$ matching has a higher large sample variance than the most efficient (or least noisy) treatment effect estimators that rely on the same assumptions. Third, \cite{AbadieImbens06} demonstrate that bootstrapping, a popular inference method based on estimating the standard error based on repeatedly resampling from the data, is inconsistent due to the discontinuous weights in pair and $1:M$ matching. The authors, however, provide a consistent asymptotic approximation of the estimator's variance based on matching within treatment groups. %to estimate the required conditional variance of $Y$ given $D$ and $X$.

To improve upon its properties, matching can be combined with a regression-based correction of the bias that stems from not fully comparable treated and nontreated matches, see \cite{Ru79} and \cite{AbIm11}. This matching-weighted regression is $n^{-1/2}$-consistent and its weights are smooth such that bootstrap inference is consistent. Another smooth method is kernel matching, which estimates $\mu_0(X)$ by a kernel function giving more weight to nontreated observations that are more similar to the treated reference observation and can attain the semiparametric efficiency bound. This requires no distance metric, as kernel functions are applied to each element in $X$ and then multiplied. Finally, genetic matching of \cite{diamond2013} matches treated and nontreated observations in a way that maximizes the balance of covariate distributions across treatment groups according to predefined balance metrics, based on an appropriately weighted distance metric.

In empirical applications, matching on the estimated propensity score is much more common than matching directly on $X$. The propensity score is typically specified parametrically by logit or probit functions. Collapsing the covariate information into a single parametric function avoids the curse of dimensionality, which implies that in finite samples, the probability of similar matches in all elements of $X$ quickly decreases in the dimension of $X$. At the same time, it allows for effect heterogeneity across $X$. On the negative side, a misspecification of the propensity score model may entail an inconsistent treatment effect estimator, which is avoided by directly matching on $X$ or using a nonparametric propensity score estimate. Matching on the estimated propensity score has a different variance than matching directly on $X$, which for the ATET can be either higher or lower, see \cite{Heck+98}. \cite{AbadieImbens2016} provide an asymptotic variance approximation for propensity score matching that appropriately accounts for uncertainty due to propensity score estimation.

Matching estimators typically require the choice of tuning parameters, be it the number of matches $M$, the bandwidth in kernel or radius matching, or the distance metric. However, theoretical guidance is frequently not available, see \cite{Froe2005} for an exception. Practitioners commonly pick tuning parameters ad hoc or based on data-driven methods that are not necessarily optimal for treatment effect estimation, as e.g.\ cross-validation for estimating $\mu_0(X)$. It appears thus advisable to investigate the sensitivity of the effect estimates w.r.t.\ varying these parameters.

As an alternative to matching, \cite{Hirano+00} discuss treatment effect estimation based on the IPW sample analog of \eqref{ipwselobs}, using series regression to obtain nonparametric plug-in estimates of the propensity score, which attains the semiparametric efficiency bounds. \cite{IchimuraLinton01} and \cite{LiRaWo04} consider IPW with kernel-based propensity score estimation. Practitioners mostly rely on logit or probit specifications, which generally is not semiparametrically efficient, see \cite{ChenHongTarozzi2008}. In any case, it is common and recommended to use normalized sample analogs of the expressions in \eqref{ipwselobs}, which ensures that the weights of observations within treatment groups sum up to one, see \cite{BuDNMC09}. Compared to matching, IPW has the advantages that it is computationally inexpensive and does not require choosing tuning parameters (other than for nonparametric propensity score estimation, if applied). On the negative side, IPW is likely sensitive to propensity scores that are very close to one or zero, see the simulations in \cite{Froe00a} and \cite{BuDNMC09} and the theoretical discussion in \cite{KhTa07}. Furthermore, IPW may be less robust to propensity score misspecification than matching, which merely uses the score to match treated and non-treated observations, rather than plugging it directly into the estimator, see \cite{Waernbaum2012}.

A variation of IPW are the empirical likelihood methods of \cite{GrahamPintoEgel2012} and \cite{ImaiRatkovic2014}. In spirit comparable to genetic matching, the methods iterate an initial propensity score estimate (e.g.\ by changing the coefficients of a logit specification) until prespecified moments of $X$ are maximally balanced across treatment groups. A related approach is entropy balancing, see \cite{hainmueller2012}, which iterates initially provided (e.g.\ uniform) weights until balance in the moments of $X$ is maximized, under the constraint that weights sum up to one in either treatment group. In contrast to methods aiming for  perfect covariate balance in prespecified moments, \cite{Zubizarreta2015} trades off balance and variance in estimation. The algorithm finds the weights of minimum variance that balance the empirical covariate distribution up to prespecified levels, i.e.\ approximately rather than exactly.

Estimation based on the sample analog of \eqref{drselobs} with plug-in estimates for $p(X), \mu_1(X), \mu_0(X)$ is called doubly robust (DR) estimation, as it is consistent if either the conditional mean outcome or the propensity score is correctly specified, see \cite{RobinsMarkNewey1992} and \cite{RoRoZa95}. If both are correctly specified, DR is semiparametrically efficient. This is also the case if the plug-in estimates are nonparametrically estimated, see \cite{Cattaneo2010}. Furthermore, \cite{RotheFirpo2013} show that nonparametric DR has a lower first order bias and second order variance than either IPW using a nonparametric propensity score or nonparametric outcome regression. This latter property is relevant in finite samples and implies that the accuracy of the DR estimator is less dependent on the accuracy of the plug-in estimates, e.g.\ the choice of the bandwidth in the kernel-based estimation of propensity scores and conditional mean outcomes. A further method satisfying the DR property is targeted maximum likelihood (TMLE), see \cite{VanderLaanRubin2006}, in which an initial regression estimate is updated (or robustified) based on an IPW parameter.

\section{Practical issues and extensions}\label{prac}

This section discusses practical issues related to propensity score methods as well as extensions of treatment evaluation to non-binary treatments and different effect parameters. One important question is whether the estimated propensity score successfully balances $X$ across treatment groups, e.g.\ in matched samples or after reweighting covariates (rather than outcomes) by IPW. Practitioners frequently consider hypothesis tests, e.g.\ two-sample t-tests applied to each element in $X$ or F-tests for jointly testing imbalances in $X$, see also the joint tests of \cite{Sianesi04} and \cite{SmTo05}. As an alternative to hypothesis tests, \cite{RosenbaumRubin1985} consider a covariate's absolute mean difference across treated and nontreated matches, divided or standardized  by the square root of half the sum of the covariate's variances in either treatment group prior to matching. In contrast to a t-test, which rejects balance under the slightest difference if the sample grows to infinity, this standardized difference is insensitive to the sample size. Rather than judging balance based on a p-value as in hypothesis tests, a standardized difference larger than a specific threshold, say 0.2, may be considered as indication for imbalance. On the negative side, the choice of the threshold appears rather arbitrary and data-driven methods for its determination are currently lacking. Taking the average of standardized differences for each covariate permits constructing a joint statistic for all covariates.

A second practical issue is whether common support in the propensity score distributions across treatment groups is sufficiently decent in the data. For the ATET, this implies that for each treated observation, nontreated matches with similar propensity scores exist, while for the ATE, this also needs to hold vice versa. Strictly speaking, common support is violated whenever for any reference observation, no observation in the other treatment group with exactly the same propensity score is available. In practice, propensity scores should be sufficiently similar, which requires defining a criterion based on which dissimilar observations may be discarded from the data to enforce common support. However, discarding observations implies that effect estimation might not be (fully) representative for the initial target population and thus sacrifices (some) external validity. On the other hand, it likely reduces estimation bias within the subpopulation satisfying common support, thus enhancing internal validity. For possible common support criteria, see for instance \cite{HeIcSmTo98}, who suggest discarding observations whose propensity scores have a density of or close to zero in (at least) one treatment group. For ATET estimation, \cite{DehejiaWahba99} propose discarding all treated observations with an estimated propensity score higher than the highest value among the nontreated. For the ATE, one additionally discards nontreated observations with a propensity score lower than the lowest value among the treated. \cite{CrumpImbensMitnik09} discuss dropping observations with propensity scores close to zero or one in a way that minimizes the variance of ATE estimation in the remaining sample. \cite{HuLeWu10} discard observations that receive a too large relative weight within their treatment group when estimating the treatment effect. See \cite{LechnerStrittmatter2019} for an overview of alternative common support criteria and an investigation of their performance in a simulation study.

The discussion so far focussed on a binary treatment, however, the framework straightforwardly extends to multivalued discrete treatments. The latter may either reflect distinct treatments (like different types of labor market programs as a job search training, a computer course, etc.) or discrete doses of a single treatment (like one, two, or three weeks of a training). Under appropriate selection-on-observable assumptions, treatment effects are identified by pairwise comparisons of each treatment value with nontreatment, or of two nonzero treatment values, if the effect of one treatment relative to the other is of interest. More formally, let $d'$ and $d''$ denote the treatment levels to be compared and $I\{A\}$ the indicator function, which is one if event $A$ holds and zero otherwise. Assume that conditions analogous to \eqref{assumpselobs} are satisfied for $D=d'$ and $D=d''$, such that conditional independence assumptions $Y(d')\bot I\{ D=d'\}|X$ and $Y(d'')\bot I\{ D=d''\}|X$ hold and the so-called generalized propensity scores satisfy the common support restrictions $\Pr(D=d'|X)>0$ and  $\Pr(D=d''|X)>0$, see \cite{Im00}. Then, replacing $D$ by $I\{D=d'\}$ and $1-D$ by $I\{D=d''\}$ as well as $p(X)=\Pr(D=1|X)$ by  $\Pr(D=d'|X)$ and $1-p(X)$ by $\Pr(D=d''|X)$ in the identification results \eqref{condmeanselobs}, \eqref{condmeanselobspscore}, \eqref{ipwselobs}, and \eqref{drselobs} yields the ATE when comparing $D=d'$ vs.\ $D=d''$ as well as the ATET when considering those with $D=d'$ as the treated. As shown in \cite{Cattaneo2010}, a range of treatment effect estimators for multivalued discrete treatments are $n^{-1/2}$-consistent and semiparametrically efficient under nonparametric estimation of the plug-in parameters. See also \cite{Le01} for a discussion of matching-based estimation with multivalued discrete treatments.

When $D$ does not have discrete probability masses but is continuously distributed, the generalized propensity score corresponds to a conditional density, denoted by $f(D=d'|X)$ to distinguish it from the previously used probability  $\Pr(D=d'|X)$. In the spirit of \eqref{condmeanselobs} for binary treatments, \cite{Flores07} proposes kernel regression of $Y$ on $D$ and $X$ for estimating the mean potential outcomes of the continuous treatment. In analogy to \eqref{condmeanselobspscore}, \cite{HiranoImbens2005} regress $Y$ on polynomials of $D$ and estimates of $f(D|X)$ along with interactions, while \cite{ImaivanDyk2004} consider subclassification by the generalized propensity score. IPW-based methods as considered in \cite{Floresetal2012} require replacing indicator functions, e.g.\ $I\{D=d'\}$, by continuous weighting functions in the identification results. Consider, for instance, the kernel weight  $K\left((D-d')/h\right)/h$, where $K$ is a symmetric second order kernel function (e.g.\ the standard normal density function) that assigns more weight to values of $D$ the closer they are to $d'$. $h$ is a bandwidth gauging by how quickly the weight decays as values in $D$ become more different to $d'$ and must go to zero as the sample size increases (albeit not too fast) for consistent estimation. Then, IPW-based identification of the ATE, for instance, corresponds to
\begin{eqnarray}
\Delta = \lim_{h \rightarrow 0} E\left[ \frac{Y\cdot K\left((D-d')/h\right)/h}{f(D=d'|X)} -  \frac{Y\cdot K\left((D-d'')/h\right)/h}{f(D=d''|X))} \right],
\end{eqnarray}
where $\lim_{h \rightarrow 0}$ means `as $h$ goes to zero'. See \cite{GalvaoWang} for a further IPW approach and \cite{Kennedyetal2017} for kernel-based DR estimation under continuous treatments, including data-driven bandwidth selection.

A further conceptual extension is the dynamic treatment framework, see for instance \cite{Ro86}, \cite{RoHeBr00}, and \cite{Lech09}. It is concerned with the evaluation of sequences of treatments (like consecutive labor market programs) based on sequential selection-on-observable assumptions w.r.t.\ each treatment. Related assumptions are also commonly imposed in causal mediation analysis aiming at disentangling a total treatment effect into various causal mechanisms, see for instance \cite{RoGr92}, \cite{Pearl01}, \cite{ImKeYa10}, \cite{TchetgenTchetgenShpitser2011}, and \cite{Huber2012}, or the survey by \cite{Huber2019}. Finally, several contributions consider effect parameters related to distributions rather than means. \cite{fir07} proposes an efficient IPW estimator of quantile treatment effects (QTE) at specific ranks (like the median) of the potential outcome distribution and derives the semiparametric efficiency bounds. \cite{DoHsu2014} suggest IPW-based estimation of the distribution functions of potential outcomes under treatment and nontreatment, see also \cite{di96} and \cite{ChFeMe13} for estimators of counterfactual distributions. \cite{Imbens03} and \cite{ImWo08} provide comprehensive reviews on treatment evaluation under selection on observables.

\section{Causal machine learning}\label{ML}

The treatment evaluation methods discussed so far consider covariates $X$ as being preselected or fixed. This assumes away uncertainty related to model selection w.r.t.\ $X$ and requires substantial or strictly speaking exact contextual knowledge about the confounders that need to be controlled for and in which functional form. In reality, however, practitioners frequently select covariates based on their predictive power for the treatment, typically without appropriately accounting for this model selection step in the causal inference to follow. Fortunately, this issue can be tackled by more recent treatment evaluation methods that incorporate machine learning to control for important confounders in a data-driven way and honestly account for model selection in the estimation process. This is particularly useful in big, and more specifically in wide (or high dimensional) data with a vast number of covariates that could potentially serve as control variables, which can render researcher-based covariate selection complicated if not infeasible.

It is important to see that when combining evaluation methods for the ATE or ATET with machine learning, henceforth called causal machine learning (CML), the data must contain sufficiently rich covariate information to satisfy the selection-on-observables assumption, just as discussed in Section \ref{selobs}. Therefore, CML is not a magic bullet that can do away with fundamental assumptions required for effect identification. However, it may be fruitfully applied if there exists a subset of covariate information that  suffices to by and large tackle confounding, but is unknown to the researcher. Under the assumption that a relative to the sample size limited subset of information permits controlling for the most important confounders, CML can be shown to be approximately unbiased, even when confounding is not perfectly controlled for.
 %Due to this property, causal machine learning will most likely revolutionize treatment evaluation, in particular as data availability keeps growing.

\cite{Chetal2018} consider for instance a CML approach called double machine learning that relies on so-called orthogonalized statistics. The latter imply that treatment effect estimation is rather insensitive to approximation errors in the estimation of $p(X), \mu_1(X), \mu_0(X)$. As discussed in Section \ref{selobs}, the sample analog of \eqref{drselobs} satisfies this (doubly) robustness property along with its desirable finite sample behaviour. In contrast, estimation based on \eqref{condmeanselobs} is rather sensitive to approximation errors of $\mu_1(X), \mu_0(X)$, while estimation based on \eqref{ipwselobs} is sensitive to errors in $p(X)$. Because DR, however, incorporates both propensity score and conditional mean outcome estimation, the approximation errors enter multiplicatively into the estimation problem, which is key for the robustness property, see for instance \cite{Farrell2015}.

A further element of many CML approaches including double machine learning is the use of independent samples for estimating the specifications of  plug-in parameters like $p(X)$, $\mu_1(X)$, and $\mu_0(X)$ on the one hand and of the treatment effects $\Delta, \Delta_{D=1}$ on the other hand. This is similar in spirit to the idea of training and testing data in conventional machine learning or cross-validation for tuning parameter selection and obtained by randomly splitting the sample. After estimating models for $p(X), \mu_1(X), \mu_0(X)$ in one part of the data, the model parameters (e.g.\ coefficients) are used in the other part to predict $p(X), \mu_1(X), \mu_0(X)$ and ultimately estimate the treatment effect. Sample-splitting prevents overfitting the models for the plug-in parameters, but comes at the cost that only part of the data are used for effect estimation, thus increasing the variance. So-called cross-fitting tackles this issue by swapping the roles of the data parts for estimating the plug-in models and the treatment effect. The treatment effect estimate is obtained as the average of the estimated treatment effects in each part and in fact, citehan just two data splits may be used for this procedure. When combining DR with sample splitting, it suffices for $n^{-1/2}$-convergence of treatment effect estimation that the estimates of $p(X), \mu_1(X), \mu_0(X)$ converge to their respective true values at a rate of $n^{-1/4}$ (or faster), see \cite{Chetal2018}. Under specific regularity conditions, this convergence rate is attained by many machine learning algorithms and even by deep learning (which is popular in computer science e.g.\ for pattern recognition), see \cite{FarrellLiangMisra2018}.

However, it needs to be stressed that CML is conceptually different to standard machine learning, which aims at accurately predicting an outcome by observed predictors based on minimizing the prediction error (e.g.\ the mean squared error) through optimally trading off prediction bias and variance. This mere forecasting approach generally does not allow learning the causal effects of any of the predictors. One reason is that a specific predictor might obtain a smaller weight (e.g.\ regression coefficient) than implied by its true causal effect if the predictor is sufficiently correlated with other predictors, such that constraining its weight hardly affects the prediction bias, while reducing the variance. Therefore, predictive machine learning with $Y$ as outcome and $D$ and $X$ as predictors generally gives a biased estimate of the causal effect of $D$, due to correlations between the treatment and the covariates. In CML, however, machine learning is not directly applied to ATE or ATET estimation, but merely for predicting the plug-in parameters, e.g.\ those of the DR expression (i.e.\ the sample analog of \eqref{drselobs}) in the case of double machine learning. To this end, three separate machine learning predictions of $D$, $Y$ among the treated, and $Y$ among the nontreated are conducted with $X$ being the predictors in each step. This is motivated by the fact that covariates $X$ merely serve the purpose of tackling confounding, while their causal effects are (contrarily to the effect of $D$) not of interest, which makes the estimation of $p(X)$, $\mu_1(X)$, and $\mu_0(X)$ a prediction problem to which machine learning can be applied.

Assume for instance that $\mu_1(X)$ and $\mu_0(X)$ are estimated by a linear lasso regression, see \cite{Tibshirani96}, where $X$ as well as higher order and interaction terms thereof may be included as predictors to allow for flexible model specifications. Including too many terms with low predictive power (as it would be the case in an overfitted polynomial regression) likely increases the variance of prediction, with little gain in terms of bias reduction. On the other hand, omitting important predictors implies a large increase in prediction bias relative to the gain in variance reduction due to a parsimonious specification. For this reason, lasso regression aims to optimally balance bias and variance through regularization, i.e.\ by shrinking the absolute coefficients obtained in a standard OLS regression towards or exactly to zero for less important predictors, e.g.\ based on cross-validation for determining the optimal amount of shrinkage. Analogously, lasso logit regression may be applied for the prediction of $p(X)$, which is a regularized version of a standard logit regression. Alternatively, lasso-based estimation of $\mu_1(X)$ and $\mu_0(X)$ can be combined with approximate covariate balancing of \cite{Zubizarreta2015} instead of estimating a propensity score model for $p(X)$, see the CML algorithm suggested by \cite{AtheyImbensWager2018}.

As discussed in \cite{Chetal2018}, lasso regression attains the required convergence rate of $n^{-1/4}$ under so-called approximate sparsity. The latter implies that the number of important covariates or interaction and higher order terms required for obtaining a sufficiently decent (albeit not perfect) approximation of the plug-in parameters is small relative to the sample size $n$. %Therefore, the condition is more likely satisfied in large samples, because more predictors are allowed to have non-zero coefficients.
To see the merits of cross-fitting, note that when disregarding the latter and instead conducting the lasso and treatment estimation steps in the same (total) data, the number of important predictors is required to be small relative to $n^{-1/2}$ rather than $n$, see \cite{Bellonietal2014}. Importantly, neither cross-fitting, nor the estimation of the plug-in parameters by some $n^{-1/4}$-consistent machine learning algorithm affects the asymptotic variance of treatment effect estimation (albeit it may matter in small samples). Therefore, CML is $n^{-1/2}$-consistent and attains the semiparametric efficiency bound as if the covariates to be controlled for in DR estimation had been correctly preselected. In large enough samples, standard errors may thus be estimated by conventional asymptotic approximations without adjustment for the machine learning steps. For a more in depth review of various machine learning algorithms and CML, see for instance \cite{AtheyImbens2019}.

\section{Effect heterogeneity, conditional effects, and policy learning}\label{MLhet}

Machine learning can also be fruitfully applied to investigate treatment effect heterogeneity across $X$, while possibly mitigating inferential multiple testing issues related to snooping for subgroups with significant(ly different) effects that might be spurious. For randomized experiments where \eqref{random} holds or under the selection-on-observables assumption \eqref{assumpselobs} with preselected $X$, \cite{AtheyImbens2016} suggest a method that builds on a modification of so-called regression trees, see \cite{Breimanetal1984}. In standard machine learning for outcome prediction, the tree structure emerges by recursively partitioning the sample with respect to the predictor space such that the sum of squared deviations of outcomes and their respective partition means is minimized. This increases outcome homogeneity within and heterogeneity between partitions. Prediction of $E[Y|X=x]$ proceeds by taking the average of $Y$ in the partition that includes the value $X=x$. This is equivalent to an OLS regression with predictors and interaction terms that are discretized according to specific threshold values in the covariate space as implied by the partitions. Cross-validation may be applied to find the optimal depth of partitions e.g.\ w.r.t.\ the mean squared error.

The causal tree approach of \cite{AtheyImbens2016} contains two key modifications when compared to standard regression trees. First, instead of $Y$, the mean difference in $Y$ across treatment groups within partitions serves as outcome in the experimental context, while under selection on observables with preselected $X$, outcomes are reweighted by the inverse of the propensity score (in analogy to \ref{ipwselobs}) prior to taking mean differences. In either case, recursive partitioning increases the homogeneity in estimated treatment effects within and its heterogeneity between partitions, in order to find the largest effect heterogeneities across subgroups defined in terms of $X$. Secondly, applying sample splitting in order to use different data parts for estimating (a) the tree's model structure and (b) the treatment effects within partitions prevents spuriously large effect heterogeneities due to overfitting.

\cite{WagerAthey2018} and \cite{AtheyTibshiraniWager2019} provide a further approach for investigating effect heterogeneity that is based on the related concept of random forests, see \cite{Breiman2001}, and also applies under selection on observables when control variables are not preselected but to be learnt from the data, see Section \ref{ML}. Random forests consist of randomly drawing many subsamples from the original data and estimating trees in each subsample. Differently to standard trees, only a random subset of predictors (rather than all) is considered at each partitioning step, which safeguards against heavily correlated trees across subsamples. Predictions are obtained by averaging over the predictions of individual trees, which makes the random forest a smooth estimator and also reduces the variance when compared to discrete partitioning of a single tree.  Forest-based predictions can therefore be represented by smooth weighting functions that bear some resemblance with kernel regression.

More concisely, the so-called generalized random forest of \cite{AtheyTibshiraniWager2019} proceeds as follows. First, both $Y$ and $D$ are predicted as a function of $X$ using random forests and leave-one-out cross-fitting. The latter implies that the outcome or treatment of each observation is predicted based on all observations in the data but its own, in order to prevent overfitting when conditioning on $X$. Second, the predictions are used for computing residuals of the outcomes and treatments, which is in the spirit of orthogonalized statistics as discussed in the context of DR in Section \ref{ML}. Third, the effect of the residuals of $D$ on the residuals of $Y$ is predicted as a function of $X$ by another random forest that averages over a large number of causal trees with residualized outcomes and treatments that use different parts of the respective subsamples for tree-modelling and treatment effect estimation. Bluntly speaking, this method combines the idea of sample splitting and orthogonalization to control for important confounders as discussed in Section \ref{ML} with the approach of \cite{AtheyImbens2016} for finding effect heterogeneity.

When comparing a single causal tree and a generalized random forest, an advantage of the former is that it directly yields an easy-to-interpret partitioning based on the most predictive covariates in terms of effect heterogeneity. On the negative side, tree structures frequently have a rather high variance such that a small change in the data may entail  quite different partitions. The generalized random forest is more attractive in terms of variance, but does not provide a single covariate partitioning due to averaging over many trees. It, however, yields an estimate of the CATE $\Delta_x=E[Y(1)-Y(0)|X=x]$, see  \eqref{cate}, such that its heterogeneity as a function of $X$ can be investigated. Also note that averaging over the estimates of $\Delta_x$ in the total sample or among the treatment provides consistent estimates of the ATE and ATET, respectively. For surveys on further machine learning methods for investigating treatment effect heterogeneity, see for instance \cite{Powersetal2018} and \cite{Knausetal2018}.

A concept related to the CATE is optimal policy learning, see e.g.\ \cite{Manski2004}, \cite{HiranoPorter2008}, \cite{Stoye2009}, \cite{QianMurphy2011}, \cite{BhattacharyaDupas2012}, and \cite{KitagawaTetenov2018}, which typically aims at optimally allocating a costly treatment in some population under budget constraints. This for instance requires analyzing which observations in terms of covariate values $X$ should be assigned the constrained treatment to maximize the average outcome. Examples include the optimal selection of jobseekers to be trained to maximize the overall employment probability or the optimal choice of customers to be offered a discount in order to maximize average sales. Formally, let $\pi'(X)$ denote a specific treatment policy defined as function of $X$. To give just one example, $\pi(X)$ could require $D=1$ for all observations whose first covariate in $X$ is larger than a particular threshold and $D=0$ otherwise. The average effect of policy $\pi'(X)$, denoted by $Q(\pi'(X))$, corresponds to the difference in mean potential outcomes under $\pi(X)$ vs.\ nontreatment of everyone:
\begin{eqnarray}\label{pollearn}
Q(\pi'(X))=E[Y(\pi'(X))-Y(0)]=E[\pi(X)\cdot \Delta_X].
\end{eqnarray}
The second equality highlights the close relationship of policy learning and CATE identification. The optimal policy, denoted by $\pi^*(X)$, maximizes the average effect among the set of all feasible policies contained in the set $\Pi$:
\begin{eqnarray}\label{pollearnopt}
\pi^*(X)=\max_{\pi\in \Pi} Q(\pi(X)).
\end{eqnarray}

\eqref{pollearn} and \eqref{pollearnopt} permit defining the so-called regret function associated with treatment policy $\pi'(X)$, which is denoted by $R\pi'(X)$ and equals the (undesirable) reduction in the average policy effect due to implementing $\pi'(X)$ rather than the optimal policy $\pi^*(X)$:
\begin{eqnarray}
R(\pi'(X))=Q(\pi^*(X))-Q(\pi'(X)).
\end{eqnarray}
Finding the optimal policy among the set of feasible policies $\Pi$, which implies that the average policy effect $Q$ is maximized and regret $R$ is equal to zero, amounts to solving the following maximization problem:
\begin{eqnarray}\label{pollearnmax}
\pi^*(X)=\max_{\pi\in \Pi} E[(2\pi(X)-1)\cdot \phi(X)].
\end{eqnarray}
Note that $\phi(X)$ is the DR statistic of \eqref{drselobs}, see for instance \cite{Dudiketal2011}, \cite{Zhangetal2012}, and \cite{Zhouetal2017} for DR-based policy learning. The term $(2\pi(X)-1)$ implies that the CATEs of treated and nontreated subjects enter positively and negatively into the expectation, respectively. Maximizing the expectation therefore requires optimally trading off treated and nontreated subjects in terms of their CATEs when choosing the treatment policy among all feasible policies. Estimation of the optimal policy may be based on the sample analog of \eqref{pollearnmax}, where $\phi(X)$ is estimated by cross-fitting and machine learning-based prediction of the plug-in parameters as outlined in Section \ref{ML}. \cite{AtheyWager2018} demonstrate that similar to ATE estimation, basing policy learning on DR machine learning has desirable properties under specific conditions, even if the important elements in $X$ driving confounding and/or effect heterogeneity are a priori unknown. The regret of the estimated optimal policy in the data when compared to the true optimal policy $\pi^*(X)$ decays at rate $n^{-1/2}$ under selection on observables if all plug-in parameters are estimated at rate $n^{-1/4}$. \cite{ZhouAtheyWager2018} show how this result extends to policy learning for multivalued discrete treatments as also considered in \cite{Kallus2017}.

%Also in the context of policy learning, the use of a DR statistic combined with cross-fitting entails desirable properties. Namely regret decays at $\sqrt{n}^{-1}^{-1}$-rate under specific conditions.

%To estimate the optimal policy in the, the sample ana and can be estimated
%\begin{eqnarray}
%\hat{\pi}=\arg\max_{\pi\in \Pi}\left\{\frac{1}{n}\sum_{i=1}^{n}(2\pi(X)-1)\cdot\hat{\phi}(X)\right\}.
%\end{eqnarray}
%$\hat{\phi}(X)$ is an estimate of $\phi(X)$ in \eqref{drselobs} which is obtained by machine learning-based estimation of the plug-in parameters as outlined in Section \ref{ML}.

\section{Instrumental variables}\label{IV}

%They represent an alternative strategy to identification based on selection on observables, see for instance the surveys in \cite{Im04} and \cite{ImWo08}, where the treatment decision is assumed to be as good as random after controlling for observed characteristics.

The selection-on-observables assumption imposed in the previous sections fails if selection into treatment is driven by unobserved factors that affect potential outcomes conditional on $X$. As an example, consider an experiment with imperfect compliance in which access to a training program is randomly assigned, but a subset of jobseekers that are offered the training does not comply and decides to not participate. If compliance behaviour is driven by unobserved factors (e.g.\ ability or motivation) that also affect the outcome (e.g.\ employment), endogeneity jeopardizes a causal analysis based on a naive comparison of treated and nontreated outcomes even when controlling for observed characteristics. However, if mere treatment assignment satisfies a so-called exclusion restriction such that it does not directly affect the outcome other than through actual treatment participation, it may serve as instrumental variable (IV), denoted by $Z$, to identify the treatment effect among those complying with the assignment. The intuition of IV-based identification is that the effect of $Z$ of $Y$, which is identified by the randomization of the instrument, only operates through the effect of $Z$ on $D$ among compliers due to the exclusion restriction. Therefore, scaling (or dividing) the average effect of $Z$ on $Y$ by the average effect of $Z$ on $D$ yields the average effect of $D$ on $Y$ among compliers, see \cite{Imbens+94} and \cite{Angrist+96}.

However, in many applications it may not appear credible that IV assumptions like random assignment hold unconditionally, i.e.\ without controlling for observed covariates. This is commonly the case in observational data in which the instrument is typically not explicitly randomized like in an experiment. For instance, \cite{Card95} considers geographic proximity to college as IV for the likely endogenous treatment education when assessing its effect on earnings. While proximity might induce some individuals to go to college who would otherwise not, e.g.\ due to housing costs associated with not living at home, it likely reflects selection into neighborhoods with a specific socio-economic status that affects labor market performance, implying that the IV is not random. If all confounders of the instrument-outcome relationship are plausibly observed in the data, IV-based estimation can be conducted conditional on observed covariates. For this reason, \cite{Card95} includes a range of control variables like parents' education, ethnicity, urbanity, and geographic region.

To formally state the IV assumptions that permit identifying causal effects conditional on covariates $X$ in the binary instrument and treatment case, denote by $D(1)$ and $D(0)$ the potential treatment decision if instrument $Z$ is set to 1 or 0, respectively. This permits defining four compliance types: Individuals satisfying $(D(1)=1,D(0)=0)$ are compliers as they only take the treatment when receiving the instrument. Non-compliers may consist of never takers who never take the treatment irrespective of the instrument $(D(1)=D(0)=0)$, always takers $(D(1)=D(0)=1)$, and defiers, who counteract instrument assignment $(D(1)=0,D(0)=1)$. Furthermore, denote (for the moment) the potential outcome as $Y(z,d)$, i.e.\ as function of both the instrument and the treatment. Then, the local average treatment effect (LATE) among compliers, denoted by $\Delta_{D(1)=1,D(0)=0}=E[Y(1)-Y(0)|D(1)=1,D(0)=0]$, is nonparametrically identified under the following assumptions, see \cite{Abadie00}.
\begin{eqnarray}\label{assumpiv}
Z \bot (D(z), Y(z',d))|X\textrm{ for }z,z',d\in\{1,0\},\quad X(1)=X(0)=X,\quad  0<P(Z=1|X)<1,\\
\Pr(D(1)\ge D(0)|X)=1,\quad  E[D|Z=1,X]-E[D|Z=0,X]\neq 0,\notag\\
\Pr(Y(1,d)=Y(0,d)=Y(d)|X)=1\textrm{ for }z,z',d\in\{1,0\}.\notag
\end{eqnarray}

The first line of \eqref{assumpiv} says that $Z$ is not deterministic in $X$ (common support) and that conditional on $X$ (which must not be affected by $D$), the IV is as good as random and thus not influenced by unobserved factors affecting the treatment and/or outcome. This is a selection-of-observables assumption similar to \eqref{assumpselobs}, however now imposed w.r.t.\ the instrument rather than the treatment. Therefore, the effects of $Z$ on $Y$ and on $D$ are identified conditional on $X$, just in analogy to the identification of the effect of $D$ on $Y$ given $X$ in Section \ref{selobs}. For this reason, replacing $D$ by $Z$ and the treatment propensity score $p(X)=\Pr(D=1|X)$ by the instrument propensity score $\Pr(Z=1|X)$ in the identification results for the ATE in \eqref{condmeanselobs}, \eqref{condmeanselobspscore}, \eqref{ipwselobs}, \eqref{drselobs} yields the average effect of the instrument on the outcome. The latter is known as intention-to-treat effect (ITT) and henceforth denoted by $\theta$. Additionally replacing $Y$ by $D$ yields the average effect of the instrument on the treatment (i.e. $E[D(1)-D(0)]$), the so-called first stage effect, denoted by $\gamma$.

The second line of \eqref{assumpiv} rules out the existence of defiers, but requires the existence of compliers conditional on $X$, due to the non-zero conditional first stage, while never and always takers might exist, too. By the law of total probability, this implies that $\gamma$ corresponds to the share of compliers, as $D(1)-D(0)$ equals one for compliers and zero for never and always takers. The third line invokes the exclusion restriction such that $Z$ must not have a direct effect on $Y$ other than through $D$. By the law of total probability, the ITT in this case corresponds to the first stage effect $\gamma$ times the LATE $\Delta_{D(1)=1,D(0)=0}$. This follows from the nonexistence of defiers and the fact that the effect of $Z$ on $Y$ is necessarily zero for always and never takers, whose $D$ is not affected by $Z$. Therefore, the LATE is identified by scaling the ITT by the first stage effect. Formally,
\begin{eqnarray}\label{LATEident}
\theta=\Delta_{D(1)=1,D(0)=0}\cdot \gamma\quad \Leftrightarrow\quad \Delta_{D(1)=1,D(0)=0}=\frac{\theta}{\gamma}.
\end{eqnarray}

If $X$ is preselected, estimation of $\Delta_{D(1)=1,D(0)=0}$ proceeds by estimating both $\theta$ and $\gamma$ based on any of the treatment effect estimators outlined in Section \ref{selobs} and by dividing one by the other, which is $n^{-1/2}$-consistent under specific regularity conditions. \cite{Froe02a}, for instance, considers nonparametric matching- and (local polynomial and series) regression-based estimation. \cite{HongNekipelov2010} derive semiparametric efficiency bounds for LATE estimation and propose efficient estimators. \cite{DoHsLi2014} and \cite{DoHsLi2014b} propose IPW estimation using series logit and local polynomial regression-based estimation of the instrument propensity score. \cite{Tan2006} and \cite{Uysal2011} discuss DR estimation with parametric plug-in parameters. If IV confounders are not preselected but in analogy to Section \ref{ML} are to be learnt from possibly high dimensional data, then causal machine learning may be applied to the DR representation of both $\theta$ and $\gamma$ in order to estimate the LATE, see for instance \cite{Bellonietal2017}. Finally, the analysis of effect heterogeneity and optimal policies discussed in Section \ref{MLhet} also extends to the IV context by using doubly robust statistics appropriate for LATE estimation, see \cite{AtheyWager2018} and \cite{AtheyTibshiraniWager2019}.

\cite{FrMe13} discuss the identification of the local quantile treatment effect on compliers (LQTE) and propose an IPW estimator based on local polynomial regression for IV propensity score estimation.  %\cite{HsLaLi15} consider series logit regression and provide estimators of the distribution and quantile functions of the complier potential outcomes.
\cite{Bellonietal2017} consider LQTE estimation based on causal machine learning when $X$ are not preselected and important instrument confounders are to be learned from the data. In contrast to the previously mentioned studies, \cite{Abadie+99} consider estimation of the conditional LQTE given particular values in $X$ by applying the so-called $\kappa$-weighting approach of \cite{Abadie00}. The latter permits identifying a broad class of complier-related statistics, based on the following weighting function $\kappa$:
\begin{eqnarray}
\kappa=1-\frac{D\cdot (1-Z)}{1-\Pr(Z=1|X)}-\frac{(1-D)\cdot Z}{\Pr(Z=1|X)}.
\end{eqnarray}
For instance, $\frac{E (\kappa \cdot X)}{E (\kappa)}=E[X|D(1)=1,D(0)=0]$ yields the mean of $X$ among compliers, which permits judging the similarity of this subgroup and the total population in terms of observed characteristics.

The LATE assumptions are partly testable by investigating specific moment inequalities w.r.t.\ outcomes across complier types that need to hold for valid instruments, see the tests proposed by \cite{Kitagawa2008}, \cite{HuMe11},  \cite{MoWa2014}, \cite{Sharma2016}, and \cite{Guber2018}. The latter uses a modified version of the causal tree of \cite{AtheyImbens2016} to increase asymptotic power by searching for the largest violations in IV validity across values $X$ in a data-driven way. %For conceptually different approaches see \cite{Slichter2014}, who tests for associations of $Z$ and $Y$ conditional on $X$ values with a zero first stage of $Z$ on $D$, and \cite{DzemskiSarnetzki2014}, who propose an overidentification test in the presence of two instruments.
It is also worth noting that even if monotonicity $\Pr(D(1)\ge D(0)|X)=1$ is violated and defiers exist, the LATE on a fraction of compliers can still be identified if a subset of compliers is equal to the defiers in terms of the average effect and population size, see \cite{deChaisemartin2016}.

When extending the binary instrument and treatment case to a multivalued instrument $Z$ and a binary $D$, LATEs are identified w.r.t.\ any pair of values $(z'',z')$ satisfying the IV assumptions. Each of them may have a different first stage and thus, complier population. Particularly interesting appears the LATE for the largest possible complier population. The latter is obtained by defining the treatment propensity score $p(z)=\Pr(D=1|Z=z,X=x)$ as instrument and considering the pair of propensity score values that maximizes compliance given $X=x$, see \cite{Froe02a}.

A continuously distributed instrument even permits identifying a continuum of complier effects under appropriately adapted IV assumptions. Specifically, a marginal change in the instrument yields the so-called marginal treatment effect (MTE), see \cite{HeckVytlacil00} and \cite{HeVy05}, which can be interpreted as the average effect among individuals who are indifferent between treatment or nontreatment given their values of $Z$ and $X$. Technically speaking, the MTE is the limit of the LATE when the change in the instrument goes to zero.

In contrast to multivalued instruments, generalizing identification from binary to nonbinary treatments is not straightforward. Assume a binary instrument and an ordered treatment  $D\in \{0, 1,...,J\}$, with $J+1$ being the number of possible (discrete) treatment doses. \cite{AngristImbens95} show that effects for single compliance types at specific treatment values, e.g.\ for those increasing the treatment from 1 to 2 when the increasing the instrument from $0$ to $1$, are not identified. It is, however, possible to obtain a non-trivially weighted average of effects of unit-level increases in the treatment on heterogeneous complier groups defined by different margins of the potential treatments. Albeit this is a proper causal parameter, its interpretability is compromised by the fact that the various complier groups generally enter with non-uniform weights. Similar issues occur if both instruments and treatments are multivalued.

There has been a controversial debate about the practical relevance of the LATE, as it only refers to the subgroup of compliers, see e.g.\ \cite{De10}, \cite{Im10}, \cite{HeUr10}. It is therefore interesting to see under which conditions this effect can be extrapolated to other populations. As discussed in \cite{Angrist2004}, the LATE is directly externally valid, i.e.,\ corresponds to the ATE when either all mean potential outcomes are homogeneous across compliance types, or at least the average effects. For testing the equality of mean potential outcomes across treated compliers and always takers as well as across nontreated compliers and never takers, see \cite{Angrist2004}, \cite{deLunaJohansson2012}, \cite{Huber2013}, and \cite{BlackJooLaLondeSmithTaylor2015}. See also \cite{DoHsLi2014} for a related, but yet different testing approach. If equality in all mean potential outcomes holds at least conditional on $X$, instruments are in fact not required for identification as selection into $D$ is on observables only, see Section \ref{selobs}. \cite{AnFe2010} and \cite{AronowCarnegie2013} do not consider homogeneity in mean potential outcomes but discuss extrapolation of the LATE when assuming homogeneous effects across compliance types. This assumption, which rules out selection into treatment by unobserved gains as assumed in standard \cite{Roy51} models, is testable if several instruments are available. For a comprehensive survey on methodological advancements in LATE evaluation, see \cite{HuberWuthrich2019}.

\section{Difference-in-Differences}\label{did}

The difference-in-differences (DiD) approach bases identification on the so-called common trend assumption. The latter says that the mean potential outcomes under nontreatment of the actually treated and nontreated groups experience a common change over time when comparing periods before and after the treatment. Assuming that both groups would in the absence of the treatment have experienced the same time trend in potential outcomes, however, permits for differences in the levels of potential outcomes due to selection bias. As an example, assume that of interest is the employment effect of a minimum wage ($D$), which is introduced in one geographic region, but not in another one, see for instance \cite{CardKrueger1994}. While the employment level ($Y$) may differ in both regions due to differences in the industry structure, DiD-based evaluation requires that employment changes e.g.\ due to business cycles would be the same in the absence of a minimum wage. In this setup, a comparison of average employment in the post-treatment period across regions does not give the effect of the minimum wage due to selection bias related to the industry structure. A before-after comparison of employment (i.e.\ before and after treatment introduction) within the treated region is biased, too, as it picks up both the treatment effect and the business cycle-related time trend. Under the common trend assumption, however, the time trend for either region is identified by the before-after comparison in the nontreated region. Subtracting the before-after difference in employment in the nontreated region (time trend) from the before-after difference in the treated region (treatment effect plus time trend) therefore gives the treatment effect on the treated. That is, taking the difference in (before-after) differences across regions yields identification under the common trend assumption.

In many empirical problems, common trends may only appear plausible after controlling for observed covariates $X$. For instance, it could be argued that the assumption is more likely satisfied for treated and nontreated subjects within the same occupation or industry. Formally, let $T$ denote a time index which is equal to zero in the pre-treatment period, when neither group received the treatment, and one in the post-treatment period, after one out of the two groups received the treatment. To distinguish the potential outcomes in terms of pre- and post-treatment periods, the subindex $t$ $\in$ $\{1,0\}$ is added, such that  $Y_0(1),Y_0(0)$ and $Y_1(1),Y_1(0)$ correspond to the pre- and post-treatment potential outcomes, respectively. The following conditions permit identifying the ATET in the post-treatment period, denoted by $\Delta_{D=1,T=1}=E[Y_1(1)-Y_1(0)|D=1,T=1]$, see the review of the DiD framework in \cite{Lechner2010}:
\begin{eqnarray}\label{didass}
E[Y_1(0)-Y_0(0)|D=1,X]&=&E[Y_1(0)-Y_0(0)|D=0,X],\quad X(1)=X(0)=X, \\
%E[Y_0(0)-Y_0(0)|D=1,X]&=&0, \quad \Pr(D=1)>0, \quad p(X)<1, \quad 0<\Pr(T=1)<1. \notag
E[Y_0(1)-Y_0(0)|D=1,X]&=&0, \notag \\
\Pr(D=1,T=1|X, (D,T)&\in&\{ (d,t),(1,1)\})<1\textrm{ for all }(d,t)\in \{(1,0),(0,1),(0,0)\}. \notag
\end{eqnarray}

The first line of \eqref{didass} imposes that $X$ is not affected by $D$ and formalizes the conditional common trend assumption stating that conditional on $X$, no unobservables jointly affect the treatment and the trend of mean potential outcomes under nontreatment. This is a selection-on-observables assumption on $D$, however, w.r.t.\ the changes in mean potential outcomes over time, rather than their levels as in \eqref{assumpselobs} of Section \ref{selobs}. The two types of assumptions are not nested, such that neither implies the other, and cannot be combined for the sake of a more general model, see the discussion in \cite{ChabeFerret2017}. The second line in \eqref{didass} rules out (average) anticipation effects among the treated, implying that $D$ must not causally influence pre-treatment outcomes in expectation of the treatment to come. The third line imposes common support: For any value of $X$ appearing in the group with $(D=1,T=1)$, subjects with such values of $X$ must also exist in the remaining three groups with $(D=1,T=0)$, $(D=0,T=1)$, and $(D=0,T=0)$.

Given that the identifying assumptions hold, the DiD strategy applies to both panel data with the same subjects in pre- and post-treatment periods as well as to repeated cross sections
with different subjects in either period. Under \eqref{didass}, $E[Y|D=0,T=1,X]-E[Y|D=0,T=0,X]=E[Y_1(0)-Y_0(0)|D=0,X]=E[Y_1(0)-Y_0(0)|D=1,X]$. This may be subtracted from $E[Y|D=1,T=1,X]-E[Y|D=1,T=0,X]=E[Y_1(1)-Y_0(1)|D=1,X]=E[Y_1(1)-Y_1(0)|D=1,X]+E[Y_1(0)-Y_0(1)|D=1,X]=E[Y_1(1)-Y_0(1)|D=1,X]=E[Y_1(1)-Y_1(0)|D=1,X]+E[Y_1(0)-Y_0(0)|D=1,X]$, where the second equality follows from subtracting and adding $Y_1(0)$ and the third from ruling out anticipation effects, in order to obtain the conditional ATET $E[Y_1(1)-Y_1(0)|D=1,X]$. Therefore, averaging over the distribution of $X$ among the treated in the post-treatment period yields the ATET in that period:
\begin{eqnarray}\label{DiDidentpost}
&&\Delta_{D=1,T=1}=E[  \mu_1(1,X)-\mu_1(0,X) - (\mu_0(1,X)-\mu_0(0,X))|D=1, T=1 ] \\
&=& E\left[ \left\{ \frac{D\cdot T}{ \Pi} - \frac{D\cdot (1-T)\cdot \rho_{1,1}(X)}{ \rho_{1,0}(X)\cdot \Pi}
- \left( \frac{(1-D)\cdot T\cdot \rho_{1,1}(X)}{ \rho_{0,1}(X) \cdot \Pi} - \frac{(1-D)\cdot (1-T)\cdot \rho_{1,1}(X)}{ \rho_{0,0}(X)\cdot \Pi}\right)\right\}\cdot Y  \right],\notag %\\
%&=& E\left[ \left\{ \frac{D\cdot T}{ \Pi} - \frac{D\cdot (1-T)\cdot %\rho_{1,1}(X)}{ \rho_{1,0}(X)\cdot \Pi}
%-  \frac{(1-D)\cdot T\cdot \rho_{1,1}(X)}{ \rho_{0,1}(X) \cdot \Pi} + %\frac{(1-D)\cdot (1-T)\cdot \rho_{1,1}(X)}{ \rho_{0,0}(X)\cdot \Pi} %\right\}\cdot (Y-\mu_0(T,X)) \right],\notag
\end{eqnarray}
where $\Pi=\Pr(D=1,T=1)$, $\rho_{d,t}(X)=\Pr(D=d,T=t|X)$, and $\mu_d(t,x)=E[Y|D=d,T=t,X=x]$.

As pointed out in \cite{Hong2013}, many DiD studies at least implicitly make the additional assumption that the joint distributions of treatment $D$ and covariates $X$ remain constant over time $T$, formalized by $(X,D)\bot T$. This for instance rules out that the composition of $X$ changes between periods in either treatment group. Under this additional assumption, $\Delta_{D=1,T=1}$ coincides with the `standard' ATET $\Delta_{D=1}$, which is then identified by the following expressions:
\begin{eqnarray}\label{DiDident}
\Delta_{D=1}&=&E[  \mu_1(1,X)-\mu_1(0,X) - (\mu_0(1,X)-\mu_0(0,X))|D=1 ] \\
%&=& E[Y|D=1,T=1]-E[Y|D=1,T=0]-E[  (\mu_0(1,X)-\mu_0(0,X))|D=1 ] \\
&=& E\left[ \left\{ \frac{D\cdot T}{ P\cdot \Lambda} - \frac{D\cdot (1-T)}{ P\cdot (1-\Lambda)}
 - \left( \frac{(1-D)\cdot T\cdot p(X)}{ (1-p(X))\cdot P\cdot \Lambda} - \frac{(1-D)\cdot (1-T)\cdot p(X)}{ (1-p(X))\cdot P\cdot (1-\Lambda)}\right)\right\}\cdot Y  \right]\notag \\
&=& E\left[ \left\{ \frac{D\cdot T}{ P\cdot \Lambda} - \frac{D\cdot (1-T)}{ P\cdot (1- \Lambda)}
 -  \left(\frac{(1-D)\cdot T\cdot p(X)}{ (1-p(X))\cdot P\cdot \Lambda} - \frac{(1-D)\cdot (1-T)\cdot p(X)}{ (1-p(X))\cdot P\cdot (1-\Lambda)}\right)  \right\}\cdot (Y-\mu_0(T,X)) \right],\notag
\end{eqnarray}
where $p(X)=\Pr(D=1|X)$, $P=\Pr(D=1)$, and $\Lambda=\Pr(T=1)$. Exploiting the identification results after the first, second, and third equalities in \eqref{DiDident}, $n^{-1/2}$-consistent estimation may be based on regression or matching, on IPW as considered in \cite{Abadie2005}, or on DR estimation as in \cite{SantAnnaZhao2018}, respectively. \cite{Zimmert2018} shows that in the presence of high dimensional covariate information, causal machine learning based on the DR representation in \eqref{DiDident} can be semiparametrically efficient in analogy to the results in Section \ref{ML}.

A general practical issue concerning DiD inference is clustering, due to a correlation in uncertainty over time (e.g.\ in panel data due to having the same subjects in either period) or within regions (e.g.\ due to being exposed to the same institutional context). In this case, observations are not independently sampled from each other, implying that inference methods not accounting for clustering might perform poorly. See e.g.\ \cite{BertrandDuflo04},  \cite{DonaldLang07}, \cite{Cameronetal2008}, \cite{ConleyTaber2011}, \cite{FermanPinto2019} for a discussion of this issue as well as of (corrections of) asymptotic or bootstrap-based inference methods under a large or small number of clusters in the treatment groups. The findings of this literature suggest that cluster- and heteroskedasticity-robust variance estimators might only work satisfactorily if the number of treated and nontreated clusters is large enough, while a small number of clusters requires more sophisticated inference methods.

The subsequent discussion reviews some methodological extensions. %, see also \cite{Lechner2010} for a comprehensive review of the DiD framework.
\cite{cha15} discuss identification when the introduction of the treatment does not induce everyone in the treatment group to be treated, but (only) increases the treatment rate citehan in the nontreated group in the spirit of an instrument, see Section \ref{IV}. \cite{AbrahamSun2018}, \cite{AtheyImbens2018DiD}, \cite{BorusyakJaravel2018}, \cite{CallawaySantAnna2018}, \cite{GoodmanBacon2018}, \cite{Hull2018}, \cite{Strezhnev2018}, \cite{cha19}, and \cite{ImaiKim2019} discuss DiD identification with multiple time periods and treatment groups that might experience treatment introduction at different points in time. \cite{Arkhangelskyetal2019} consider unit- and time-weighted DiD estimation.

\cite{AtheyImbens06} suggest the so-called Changes-in-Changes (CiC) approach, which is related to DiD in that it exploits differences in pre- and post-treatment outcomes, however, based on different (and non-nested) identifying assumptions. While CiC does not invoke any common trend assumption, it imposes that potential outcomes under nontreatment are strictly monotonic in unobserved heterogeneity and that the distribution of the latter remains constant over time within treatment groups. Such a conditional independence between unobserved heterogeneity and time is satisfied if the subjects' ranks in the outcome distributions within treatment groups do not systematically change from pre- to post-treatment periods. In contrast to DiD, CiC allows identifying both the ATET and QTET, but generally requires a continuously distributed outcome for point identification.

Finally, another approach related to, but in terms of identification yet different from DiD is the synthetic control method of \cite{AbadieGardeazabal2003} and \cite{Abadieetal2010}, which was originally developed for case study set ups with only one treated, but many nontreated units. It is based on appropriately weighting nontreated units to synthetically impute the treated unit's potential outcome under nontreatment. See e.g.\ the review article of \cite{AbadieCattaneo2018} which contains a section on the synthetic control method that provides references to methodological advancements.

\section{Regression discontinuity and kink designs}\label{rdd}

The regression discontinuity design (RDD), see \cite{Thistlethwaite60}, is based on the assumption that at a particular threshold of some observed running variable, the treatment status either changes from zero to one for everyone (sharp design) or for a subpopulation (fuzzy design). As an example, assume that the treatment of interest is extended eligibility to unemployment benefits, to which only individuals aged 50 or older are entitled, see for instance \cite{Lalive2008785}. The idea is to compare the outcomes (like unemployment duration) of treated and nontreated subjects close to the (age) threshold, e.g.\ of individuals aged 50 and 49, who are arguably similar in characteristics potentially affecting the outcome, due to their minor difference in age. The RDD therefore aims at imitating the experimental context at the threshold to evaluate the treatment effect locally for the subpopulation at the threshold.

Formally, let $R$ denote the running variable and $r_0$ the threshold value. If the treatment is deterministic in $R$ such that it is one whenever the threshold is reached or exceeded, i.e.\ $D=I\{R\geq r_{0}\}$, the RDD is sharp: All individuals change their treatment status exactly at $r_{0}$. Identification in the sharp RDD relies on the assumption that mean potential outcomes $E[Y(1)|R]$ and $E[Y(0)|R]$ are continuous and sufficiently smooth around $R=r_0$, see e.g.\ \cite{HahnToodKlaauw01}, \cite{Porter03}, and \cite{Lee08}, meaning that any factors other than $D$ that affect the outcome are continuous at the threshold. Continuity implies that if treated and nontreated populations with values of $R$ exactly equal to $r_0$ existed, the treatment would be as good as randomly assigned w.r.t.\ mean potential outcomes. This corresponds to a local selection-on-observables assumption conditional on $R=r_0$. Furthermore, the density of the running variable $R$ must be continuous and bounded away from zero around the threshold, such that treated and nontreated observations are observed close to $R=r_0$.

Under these assumptions, the ATE at the threshold, denoted by $\Delta_{R=r_0}$, is identified based on treated and nontreated outcomes in a neighbourhood $\varepsilon>0$ around the threshold when letting $\varepsilon$ go to zero:
\begin{eqnarray}\label{sharp}
&&\underset{\varepsilon \rightarrow 0}{\lim }E[ Y| R\in [r_{0}, r_{0}+\varepsilon) ]-\underset{\varepsilon \rightarrow 0}{\lim }E[ Y| R\in [r_{0}-\varepsilon, r_{0}) ]\\
&=&\underset{\varepsilon \rightarrow 0}{\lim }E[ Y(1)|R\in [r_{0}, r_{0}+\varepsilon) ]-\underset{\varepsilon \rightarrow 0}{\lim }E[ Y(0)|R \in [r_{0}-\varepsilon, r_{0}) ]=E[Y(1)-Y(0)|R=r_{0}]=\Delta_{R=r_0}.\notag
\end{eqnarray}

In the fuzzy RDD, $D$ is not deterministic in $R$ but may also depend on other factors. It is, however, assumed that the treatment share changes discontinuously at the threshold. Assume e.g.\ that admittance to a college ($D$) depends on passing a particular threshold of the score in a college entrance exam ($R$). While some students might decide not to attend college even if succeeding in the exam, a discontinuous change in the treatment share occurs if compliers exists that are induced to go to college when passing the threshold. Denote by $D(z)$ the potential treatment state as a function of the binary indicator $Z=I\{R\geq r_{0}\}$, which serves as instrument in an analogous way as discussed in Section \ref{IV}. Similar to \cite{Dong2014}, assume that around the threshold, defiers do not exist and that the shares of compliers, always takers, and never takers as well as their mean potential outcomes under treatment and nontreatment are continuous. This implies that IV-type assumptions similar to those postulated in $\eqref{assumpiv}$ conditional on $X$ hold conditional on $R=r_0$.

Under these conditions, the first stage effect of $Z$ on $D$, denoted by $\gamma_{R=r_0}$ is identified by
\begin{eqnarray}
&&\underset{\varepsilon \rightarrow 0}{\lim }E[ D| R\in [r_{0}, r_{0}+\varepsilon) ]-\underset{\varepsilon \rightarrow 0}{\lim }E[ D| R\in [r_{0}-\varepsilon, r_{0}) ]\\
&=&\underset{\varepsilon \rightarrow 0}{\lim }E[ D(1)|R\in [r_{0}, r_{0}+\varepsilon) ]-\underset{\varepsilon \rightarrow 0}{\lim }E[ D(0)|R \in [r_{0}-\varepsilon, r_{0}) ]=E[D(1)-D(0)|R=r_{0}]=\gamma_{R=r_0}.\notag
\end{eqnarray}
Furthermore, the first line of \eqref{sharp} identifies the ITT effect of $Z$ on $Y$ at the threshold, denoted by $\theta_{R=r_0}$ in the fuzzy RDD (rather than $\Delta_{R=r_0}$ as in the sharp RDD). In analogy to \eqref{LATEident} in Section \ref{IV}, the LATE on compliers at the treshold, denoted by  $\Delta_{D(1)=1,D(0)=0,R=r_0}=E[Y(1)-Y(0)|D(1)=1,D(0)=0, R=r_{0}]$, is identified by dividing the ITT by the first stage effect at the threshold:
\begin{eqnarray}
\Delta_{D(1)=1,D(0)=0,R=r_0}=\frac{\theta_{R=r_0}}{\gamma_{R=r_0}}
\end{eqnarray}

In empirical applications of the RDD, the treatment effect is predominantly estimated by a local regression around the threshold. Practitioners for instance frequently use a linear regression for estimating $E[Y|D=0,R<r_0]$ and $E[Y|D=1,R\geq 0]$ within some bandwidth around $r_0$ in order to estimate $\Delta_{R=r_0}$  by the difference of the regression functions at $r_0$ in the case of the sharp RDD. A smaller bandwidth decreases estimation bias, because observations closer to the threshold are more comparable and effect estimation is more robust to model misspecification, see \cite{Imbens18}, but increases the variance due to relying on a lower number of observations. \cite{Imbens01072012} propose a method for bandwidth selection that minimizes the squared error of the estimator. However, the optimal bandwidth for point estimation is generally suboptimal (and too large) for conducting inference, e.g.\ for computing confidence intervals. For this reason, \cite{ECTA1465} propose inference methods that are more robust to bandwidth choice and yield confidence intervals more closely matching nominal coverage, along with optimal bandwidth selection for inference. Their results imply that when $\Delta_{R=r_0}$ is estimated by linear regression within some bandwidth, then quadratic regression (i.e.\ one order higher) with the same bandwidth should be used for the computation of the standard error and confidence intervals. \cite{ArmstrongKolesar2018} suggest an alternative approach to inference that takes into account the worst case bias that could arise given a particular bandwidth choice. \cite{CattaneoFrandsenTitiunik2015} develop randomization methods for exact finite sample inference in the RDD under somewhat stronger identifying assumptions.

The identifying assumptions of the RDD are partly testable in the data. \cite{Mccrary08} proposes a test for the continuity or the running variable at the threshold, as a discontinuity points to a manipulation of $R$ and selective bunching at a one side of the threshold. In the previous example based on \cite{Lalive2008785}, certain employees and companies might for instance manipulate age at entry into unemployment by postponing layoffs such that the age requirement for extended unemployment benefits is just satisfied. As a further test, \cite{Lee08} suggests investigating whether observed pre-treatment covariates $X$ are locally balanced at either side of the threshold. Covariates also permit weakening the RDD assumptions to only hold conditional on $X$, implying that all variables jointly affecting manipulation at the threshold and the outcome are observed, see \cite{FrHu2018} who propose a nonparametric kernel estimator in this context. In contrast, \cite{CalonicoCattaneoTitiunik2016} do not exploit covariates for identification, but investigate variance reductions when linearly controlling for $X$ and provide methods for optimal bandwidth selection and robust inference for this case.

Several studies investigate conditions under which the rather local RDD effect can be extrapolated to other populations. \cite{DongLewbel2011} show the identification of the derivative of the RDD treatment effect in both sharp and fuzzy designs, which permits identifying the change in the treatment effect resulting from a marginal change in the threshold.  \cite{AngristRokkanen2015} test whether the running variable's association with the outcome vanishes on either side of the threshold conditional on covariates $X$. For the case of the sharp RDD, this implies that $X$ is sufficient to control for confounding just as under the selection-on-observables framework of Section \ref{selobs}, such that effects are also identified away from the threshold. In context of the fuzzy RDD, \cite{BertanhaImbens2019} propose a test for the equality in mean outcomes of treated compliers and always takers, as well as of untreated compliers and never takers. This permits investigating whether the effect on compliers at the threshold may be extrapolated to all compliance types at and away from the threshold. \cite{Cattaneoetal2019} demonstrate extrapolation under multiple thresholds, i.e.\ when the threshold may vary for various subjects instead of being equal for everyone, as considered in \cite{Cattaneoetal2016}.

\cite{LeeCard08}, \cite{Dong2015}, \cite{KolesarRothe2018} discuss identification and inference when the forcing variable is discrete rather than continuous, which is highly relevant for empirical applications. \cite{Papayetal2011} and \cite{KeeleTitiunik2015} extend the regression-discontinuity approach to multiple running variables. \cite{ImbensWager2019} propose an optimization-based inference method for deriving the minimax linear RDD estimator which can be applied to continuous, discrete, and multiple running variables. \cite{FrFrMe2012} discuss the identification of quantile treatment effects in the RDD. See also \cite{ImbensLemieux08} and \cite{LeLe09} for surveys on the applied and theoretical RDD literature.

Related to the fuzzy RDD is the regression kink design (RKD), see \cite{Cardetal2015}, which is technically speaking a first derivative version of the former. The treatment is assumed to be a continuous function of the running variable $R$ (rather than discontinuous as in the RDD), with a kink at $r_0$. This implies that the first derivative of $D$ w.r.t.\ $R$ (rather than the level of $D$ as in the RDD) is discontinuous at the threshold. In \cite{Landais2015}, for instance, unemployment benefits ($D$) are a kinked function of the previous wage ($R$): $D$ corresponds to $R$ times a constant percentage up to a maximum previous wage $r_0$ beyond which $D$ does not increase any further but remains constant. For this piecewise linear function, the derivative of $D$ w.r.t.\ $R$ corresponds to the percentage for $R<r_0$ and to zero for $R\geq 0$. As the treatment is deterministic in the running variable, this is known as sharp RKD.

Given appropriate continuity and smoothness conditions w.r.t.\ mean potential outcomes and the density of $R$ around $r_0$, scaling the change in the first derivatives of mean outcomes w.r.t.\ to $R$ at the threshold by the corresponding change in first derivatives of $D$ identifies a causal effect. The latter corresponds to the average derivative of the potential outcome with respect to $D$ when the latter corresponds to its value at the threshold, denoted by $d_0$, within the local population at $R=r_0$:
  \begin{eqnarray}\label{rkd}
\Delta_{R=r_0}(d_0)=\frac{\partial E[Y(d_0)|R=r_0]}{\partial D}=  \frac{\underset{\varepsilon \rightarrow 0}{\lim } \frac{\partial E[ Y| R\in [r_{0}, r_{0}+\varepsilon) ]}{\partial R} -
\underset{\varepsilon \rightarrow 0}{\lim } \frac{\partial E[ Y| R\in [r_{0}-\varepsilon, r_{0}) ]}{\partial R}
 }{  \underset{\varepsilon \rightarrow 0}{\lim }  \frac{\partial D| R\in [r_{0}, r_{0}+\varepsilon) }{\partial R} -  \underset{\varepsilon \rightarrow 0}{\lim } \frac{ \partial D| R\in [r_{0}-\varepsilon, r_{0}) }{\partial R} }
\end{eqnarray}

The fuzzy RKD permits deviations from the kinked function characterizing how the running variable affects the treatment, such that $D$ is not deterministic in $R$, see for instance \cite{SimonsenSkipperSkipper2016} for a study investigating the price sensitivity of product demand. Under specific continuity conditions and the monotonicity-type assumption that the kink of any individual either goes in the same direction or is zero, a causal effect at the threshold is identified among individuals with nonzero kinks. To this end, the derivatives of the treatment in \eqref{rkd}, namely $\frac{\partial D| R\in [r_{0}, r_{0}+\varepsilon) }{\partial R}$ and $\frac{ \partial D| R\in [r_{0}-\varepsilon, r_{0}) }{\partial R}$, are to be replaced by the derivatives of expectations
$\frac{\partial E[ D| R\in [r_{0}, r_{0}+\varepsilon) ]}{\partial R}$ and $\frac{\partial E[ D| R\in [r_{0}-\varepsilon, r_{0}) ]}{\partial R}$. As the expectation of a treatment maybe continuous even if the treatment itself is not, the fuzzy RKD may also be applied to a binary $D$, see \cite{Dong2014}.  \cite{ECTA1465} provide robust inference methods for the RKD, while \cite{GaJa2018} propose a permutation method for exact finite sample inference.

\section{Conclusion}\label{conclusion}

This chapter provided an overview of different approaches to policy evaluation for assessing the causal effect of a treatment on an outcome. Starting with an introduction to causality and the experimental evaluation of a randomized treatment, it subsequently discussed identification and flexible estimation under selection on observables, instrumental variables, difference-in-differences, changes-in-changes, and regression discontinuities and kinks. Particular attention was devoted to  approaches combining policy evaluation with machine learning to provide data-driven procedures for tackling confounding related to observed covariates, investigating effect heterogeneities across subgroups, and learning optimal treatment policies. In a world with ever increasing data availability, such causal machine learning methods aimed at optimally exploiting large amounts of information for causal inference will likely leverage the scope of policy evaluation to unprecedented levels. Besides the classic domain of public policies, this concerns not least the private sector, with ever more firms investing in data analytics to assess and optimize the causal impact of their actions like price policies or advertising campaigns.

\setlength\baselineskip{14.0pt}
\bibliographystyle{agsm}
{\footnotesize
\bibliography{research}
}

\end{document}